\documentstyle[12pt,epsf]{article}
\def\sublabel#1#2{\@bsphack\if@filesw {\let\thepage\relax
   \def\protect{\noexpand\noexpand\noexpand}%
   \edef\@tempa{\write\@auxout{\string
      \newlabel{#1}{{\@currentlabel#2}{\thepage}}}}%
   \expandafter}\@tempa
   \if@nobreak \ifvmode\nobreak\fi\fi\fi\@esphack}

\newcommand{\twographs}[2]{%
   \unitlength=1in
   \begin{center}
     \begin{picture}(6,2.2)
       \put(0,0){\labgraph{a}{#1}}
       \put(3,0){\labgraph{b}{#2}}
     \end{picture}
   \end{center}}

\newcommand{\labgraph}[2]{%
   \begin{picture}(3,2)
       \put(0,0){\makebox(3,2)%
{\centering\epsfxsize=2.6in\leavevmode\epsffile{#2.ps}}}
       \put(0,1){\makebox(0,0)[l]{(#1)}}
   \end{picture}}

\begin{document}

\baselineskip 18pt

\newcommand{\sheptitle}
{Infra-red fixed point structure characterising SUSY SU(5) symmetry breaking}

\newcommand{\shepauthor}
{B. C. Allanach$^1_a$, G. Amelino-Camelia$^2_b$ and
O. Philipsen$^2_c$}

\newcommand{\shepaddress}
{1. Department of Particle Physics, Rutherford Appleton Laboratory,
Chilton, Didcot, Oxon, OX11 OQX, UK \\
2. Theoretical Physics, University of Oxford, 
1 Keble Road, Oxford OX1 3NP, UK}

\newcommand{\shepabstract}
{We analyze the 
one-loop renormalisation group equations
for the parameters of the Higgs
potential of a supersymmetric SU(5) model 
with first step of symmetry breaking
involving an adjoint Higgs.
In particular, we investigate
the running  
of the parameters that decide the first step
of symmetry breaking in an attempt to
establish which symmetry-breaking scenarios would be most likely
if the model is the effective low-energy description
of some more fundamental theory.
An infra-red fixed point is identified analytically. 
We show that it is located at the boundary
between the region of Higgs parameter space
corresponding to unbroken SU(5) and the region corresponding
to the breaking of SU(5) to the Standard Model,
and we elaborate on its implications.
We also observe that certain forms of the Higgs potential
discussed at tree level in the 
literature are not 
renormalisation group invariant.}

\begin{titlepage}
\begin{flushright}
{\tt 
hep-ph/9611286}\\
OUTP-96-63P\\
RAL-TR-96-090\\
\end{flushright}
\vspace{.4in}
\begin{center}
{\large{\bf \sheptitle}}
\bigskip \\ \shepauthor \\ \mbox{} \\ {\it \shepaddress} \\
\vspace{.5in}
{\bf Abstract} \bigskip \end{center} \setcounter{page}{0}
\shepabstract
\vfill \noindent
{\tt a) bca@hep.phys.soton.ac.uk \\
b) gac@thphys.ox.ac.uk \\
c) owe@thphys.ox.ac.uk \\ }
\end{titlepage}

One of the non-predictive aspects of GUTs (Grand Unification Theories)
is the SSB (Spontaneous Symmetry Breaking) pattern~\cite{SSBGUTs}.
Even after selecting the matter (Higgs) content, most GUTs
may break in several different ways depending on which component
of the Higgs field acquires a vacuum expectation value, 
and this in turn depends on
the (free input) parameters in the Higgs potential of the model.

In this paper,
we consider 
the GUT as a low-energy effective
description of some more fundamental theory\cite{ross,gacrus}, 
possibly including
gravity, and therefore the parameters of the Higgs potential
are assumed to become meaningful at some scale $M^*$
(possibly given by the Planck scale $M_P \! \sim \! 10^{19} GeV$)
higher than the GUT scale $M_X$
(the scale,
of order $10^{16}$~GeV,
where the low energy couplings unify).
{}From this viewpoint it makes sense to study the RGEs (Renormalisation
Group Equations) describing the running of the parameters of the Higgs
potential between $M^*$ and $M_X$; in fact,
if strong infra-red structures were encountered in these RGEs,
it could then be argued that some SSB directions are
more {\it natural}
\/than others.
For example, a given direction of SSB would be considered to be natural
if a strongly attractive infra-red fixed point
was found within the corresponding region of Higgs parameter space,
since then values for the parameters of the Higgs potential
corresponding to the given direction of SSB
could be obtained at the GUT scale from rather generic input
values at the scale $M^*$.

This viewpoint is related to the one adopted in the
recent literature\cite{ross} in which
predictions for the low-energy values of certain
quantities are
obtained from the infra-red structure
of the relevant RGEs.
The results of those investigations
lead to the observation that
the values of the
(low-energy) parameters 
relevant 
for the description of the known physics 
are strongly influenced by the infra-red structure of RGEs.
This encourages an 
attempt to ``understand''
the SSB pattern as a possible result of renormalisation group flow.
In this letter, in order to illustrate this idea and test its viability,
we analyze the first
(GUT-scale) step of SSB
in a SUSY (supersymmetric) SU(5) GUT, 
which involves the Higgs of the 24-dimensional
irreducible representation (the adjoint).
Besides the $\underline{24}$, 
the Higgs sector of the minimal  
SUSY SU(5) model also 
includes $\underline{5} + \overline{\underline{5}}$
Higgs, which are used in the second SSB step.
However, for simplicity 
in our analysis 
of the first SSB step 
we neglect the effects of the $\underline{5} +
\overline{\underline{5}}$ Higgs. 
We therefore limit our analysis to the potentials involving
the $\underline{24}$ Higgs.
The superpotential is taken to be~\cite{su5}
\begin{equation}
W = \lambda_1 \mbox{Tr}(\Sigma^3) + \mu \mbox{Tr} (\Sigma^2) \, ,
\label{Wsu5}
\end{equation}
where $\Sigma$ denotes the $24$-dimensional superfield multiplet.
We assume that SUSY breaking is explicit,
via the ``soft'' SUSY-breaking terms in the potential
\begin{equation}
V_{soft} = \left[\frac{m_3}{6} \mbox{Tr} (\sigma^3) + m_2^2 \mbox{Tr}
(\sigma^2) +\frac{M}{2} \lambda \lambda + {\rm h.c.}\right]
+ m^2_{3/2} \mbox{Tr}(\sigma^{\dag}\sigma) \;,
\label{su5soft}
\end{equation}
where $\sigma$ represents the scalar component of $\Sigma$ and $\lambda$
denotes the SU(5) gaugino. 
The full
Higgs potential relevant 
for the first step of SSB
can be written as
\begin{eqnarray}
V = \left | {\partial W \over \partial \Sigma_i} \right |^2
+ V_{soft} + \mbox{D-terms}~.\label{pot}
\end{eqnarray}
Based on hierarchy arguments~\cite{hier} 
we expect $m_3, m_{3/2}, M \sim 1$~TeV
and $m_2 \sim 10^{11}$~GeV,
while $\mu$ is a GUT scale parameter expected
to be of order 10$^{16}$~GeV.
In the special case in which $V_{soft}$
results from simple models of spontaneously broken supergravity 
\cite{bar82} the soft breaking
parameters at tree level are constrained by
\begin{equation} \label{const}
m_3=6 m_{3/2} A \lambda_1\;,\quad m_2^2= (A-1) m_{3/2} \mu\;,
\end{equation}
with only one free parameter $A$ for the bi- and trilinear terms.
In this hypothesis,
the quantity $\delta_{3/2} \! \equiv \! m_{3/2}/\mu$
measures the relative strength of SUSY breaking in units of $M_{GUT}$.
For $\delta_{3/2} \! = \! 0$,
the scalar
potential has three degenerate minima with invariances SU(5),
SU(4)$\otimes$U(1) and $G_{SM}$~$\equiv$~SU(3)$\otimes$SU(2)$\otimes$U(1)
(the Standard Model gauge group). 
For the phenomenologically
relevant
case $|\delta_{3/2}| \! \ll \! 1$,
one can simply examine the 
corrections to the scalar potential of first order in 
$\delta_{3/2}$ that split the degeneracy \cite{su5},
\begin{equation}
V_{soft} = \frac{8 \, \mu^4 \, \delta_{3/2}}{27 \, \lambda_1^2} \, 
(A-3) \, b \, + O(\delta_{3/2}^2),
\end{equation}
where $b \! = \! 30$ for the $G_{SM}$-invariant
minimum, and $b \! = \! 20/9$
for the SU(4)$\times$U(1)-invariant minimum. 
(Obviously, $b \! = \! 0$ in the minimum preserving
the full SU(5) invariance.)
The direction of SU(5) breaking determined
by the vacuum expectation value~$\langle \sigma \rangle$ 
can then be read off
the parameter $A$.  
The case $A > 3$ does not reproduce the Standard Model phenomenology
since then the absolute minimum corresponds to unbroken SU(5)
(and even the SU(4)$\otimes$U(1)-invariant 
minimum is energetically lower than the
$G_{SM}$-invariant one).
On the other hand, for $A<3$ the lowest minimum of the potential is
$G_{SM}$-invariant (while the SU(4)$\otimes$U(1)-invariant 
minimum is energetically lower than the
SU(5)-invariant one), 
leading to the phenomenologically
plausible scenario of SU(5) breaking to $G_{SM}$ at the GUT scale.

This concludes the tree-level analysis. 
It appears quite satisfactory that the phenomenologically
plausible scenario simply requires $A<3$, which would seem
to correspond (assuming a simple-minded measure) to roughly
half of the parameter space. However, from the 
point of view advocated here,
one would like to check whether phenomenologically
plausible scenarios follow from rather generic
choices of input parameter at the scale (higher than the GUT scale)
where the GUT becomes meaningful 
as an effective low-energy description.
Let us therefore consider the running of the 
parameters of the Higgs potential.
The one-loop RGEs 
may be easily derived\footnote{In our 
calculation we take the matrix representation of the chiral superfield 
to be $\Sigma =T^a\Phi^a$, 
where the generators $T^a$ of the fundamental
representation of SU(5) are normalised by
Tr$(T^a T^b) \! = \! \delta^{ab} / 2$.
Some of the RGEs (\ref{su5RGEsa}-\ref{su5RGEsg}) 
have been previously derived 
in ref.~\cite{polpom} using a different field normalisation.}
following the general prescriptions of Martin and Vaughn~\cite{MandV}, 
\begin{eqnarray}
16 \pi^2 \frac{d \lambda_1}{d t} &=& 
3 \lambda_1 \left(\frac{189}{40} \lambda_1^2 - 10 g^2\right) 
\label{su5RGEsa} \\
16 \pi^2 \frac{d \mu}{d t} &=& 
2 \mu \left(\frac{189}{40} \lambda_1^2 - 10 g^2\right)
\label{su5RGEsb} \\
16 \pi^2 \frac{d m_3}{d t} &=& 	
3 \left[m_3\left(\frac{189}{40} \lambda_1^2 - 10 g^2\right)
+\frac{189}{20}\lambda_1^2
m_3 + 120 M\lambda_1 g^2 \right] 
\label{su5RGEsc} \\
16 \pi^2 \frac{d m_2^2}{d t} &=&
2 \left[m_2^2 \left(\frac{189}{40} \lambda_1^2 - 10 g^2\right) +
\frac{63}{40}\lambda_1\mu m_3 + 20 M \mu g^2 \right]
\label{su5RGEsd} \\
16\pi^2 \frac{d m^2_{3/2}}{d t} &=&
\frac{567}{20}\lambda_1^2 m_{3/2}^2 + \frac{21}{80} m_3^2 - 40 M^{\dag}Mg^2
\label{su5RGEse} \\
16 \pi^2 \frac{d g^2}{d t} &=&
\beta g^4 
\label{su5RGEsf} \\
16 \pi^2 \frac{d M}{d t} &=&
\beta g^2 M,
\label{su5RGEsg}
\end{eqnarray}
where $t = \ln(q^2/M_X^2)$, $q$ is the $\overline{MS}$ renormalisation
scale and the one loop beta function, $\beta = 2(S(R)-15)$,
is determined by the sum over all the Dynkin indices
of the fields in the theory, $S(R)$.
$\beta \! = \! -8$
for our SUSY SU(5) model, which hosts the above mentioned Higgs
sector plus
$3(\underline{10} \oplus \overline{\underline{5}})$
representations corresponding to 3 Standard Model fermionic families
(and superpartners).

The different evolution of $m_3$, $m_2^2$ and $m_{3/2}$ implies
that the constrained parameterisation (\ref{const}) is not 
renormalisation group invariant
and consequently the above tree-level analysis of
symmetry breaking 
is not sufficient.
In generalising the analysis of symmetry breaking
to the case of running parameters in 
the full potential (\ref{pot}), it is 
appropriate to consider the three independent parameters
$\delta_2 \! \equiv \! m^2_2/\mu^2$,
$\delta_3 \! \equiv \! m_3/\mu$ and 
$\delta_{3/2} \! \equiv \! m_{3/2}/\mu$.
In terms of these parameters, the soft-breaking potential 
can be written as\footnote{An interesting alternative to the conventional
scenario that we consider is the one of ``radiative breaking'' at the
GUT scale. 
In particular, this would require considering in what follows
the possibility $\mu \! = \! 0$, which is stable under the
one-loop RGEs.
In the present work we shall ignore this possibility. 
Its analysis would require a generalisation of our study of the
Higgs potential, not relying on the simplifications 
we achieved by assuming
$|m_i/\mu| \! \ll \! 1$.}
\begin{equation} \label{delv}
V_{soft}=\frac{8\mu^4}{27\lambda_1^2} \, b \, F \, ,
\end{equation}
where
\begin{equation} \label{newvsoft}
F \equiv 3\delta_2-\frac{1}{3\lambda_1}\delta_3
+\frac{3}{2}\delta_{3/2}^2 \,~.
\end{equation}
Hence, for $F \! < \! 0$ the $G_{SM}$-invariant minimum
is the lowest one, while SU(5) will remain unbroken
for $F \! > \! 0$.
The value of $F$ at the GUT scale $M_X$ 
determines the type of 
residual symmetry
below $M_X$.
 
To render the fixed point structure explicit, 
from (\ref{su5RGEsa}-\ref{su5RGEsg}) 
we form the following RGEs for dimensionless ratios 
\begin{eqnarray}
16 \pi^2 \frac{d}{d t} \left( \frac{\lambda_1^2}{g^2}
\right) &=& 
6 g^2 \left( \frac{\lambda_1^2}{g^2} \right) \left[ \frac{189}{40} \left(
\frac{\lambda_1^2}{g^2} \right)  - 10 - \frac{\beta}{6} \right]
\label{FP1} \\
16 \pi^2 \frac{d}{d t} \left( \frac{m_3}{M \lambda_1}
\right) &=& 9 g^2 \left[  \left( \frac{m_3}{M \lambda_1}
\right) \left[ \left( \frac{\lambda_1^2}{g^2}
\right)  \frac{63}{20} - \frac{\beta}{9} \right] + 40 \right]
\label{FP2} \\
16 \pi^2 \frac{d}{d t} \left( \frac{m_2^2}{M \mu}
\right) &=&   g^2 \left[ - \beta \left( \frac{m_2^2}{M \mu}
\right) +  \frac{63}{20} \left( \frac{m_3}{M \lambda_1}
\right) \left( \frac{\lambda_1^2}{g^2}
\right) + 40 \right]~.
\label{FP3}
\end{eqnarray}
The right hand side of this system of coupled equations vanishes
for 
\begin{equation}
\left( \frac{\lambda_1^2}{g^2}
\right)^* = \frac{40}{189} (10 + \beta / 6) , 
\quad
\left( \frac{m_3}{M \lambda_1}
\right)^* = -6, \label{secFP}
\quad
\left( \frac{m_2^2}{M \mu}
\right)^* = - \frac{2}{3}. \label{thFP}
\end{equation}
The fixed point described by Eq.~(\ref{thFP}) is a specific example of a
more general class of fixed points identified in ref.~\cite{JJ}.
By linearising (\ref{FP1}-\ref{FP3})
around the fixed point one easily finds that
it is infra-red stable 
when $\beta \! < \! 0$,
as in the case 
of the SUSY SU(5) model considered here.
For $\beta \! > \! 0$, which can be achieved
by adding more matter to the model, one would have
a saddle point.
Assuming $\delta_{3/2} \! \ll \! 1$, as implied by hierarchy arguments, 
we may neglect the 
second order contribution of order $\delta_{3/2}^2$, and $F$ is well
approximated by 
\begin{equation} \label{fapp}
F\approx 3\delta_2-\frac{1}{3\lambda_1}\delta_3
=\frac{M}{\mu} \left[ 3 \frac{m_2^2}{M\mu}
-\frac{1}{3} \frac{m_3}{M\lambda_1} \right]\;,
\end{equation}
which is zero at the fixed point.
Thus, starting at some scale $M^*$, {\it e.g.}
\/the Planck scale, and running to the GUT scale, 
the Higgs parameters evolve towards values at
the boundary ($F \! = \! 0$) between the 
region of parameter space corresponding to unbroken SU(5)
and the region of parameter space corresponding
to SU(5) breaking to $G_{SM}$\footnote{If there are significant
contributions from $\delta_{3/2}^2$
the flow to the unbroken-SU(5) region
is favoured.}.

\begin{figure}[ht]
\twographs{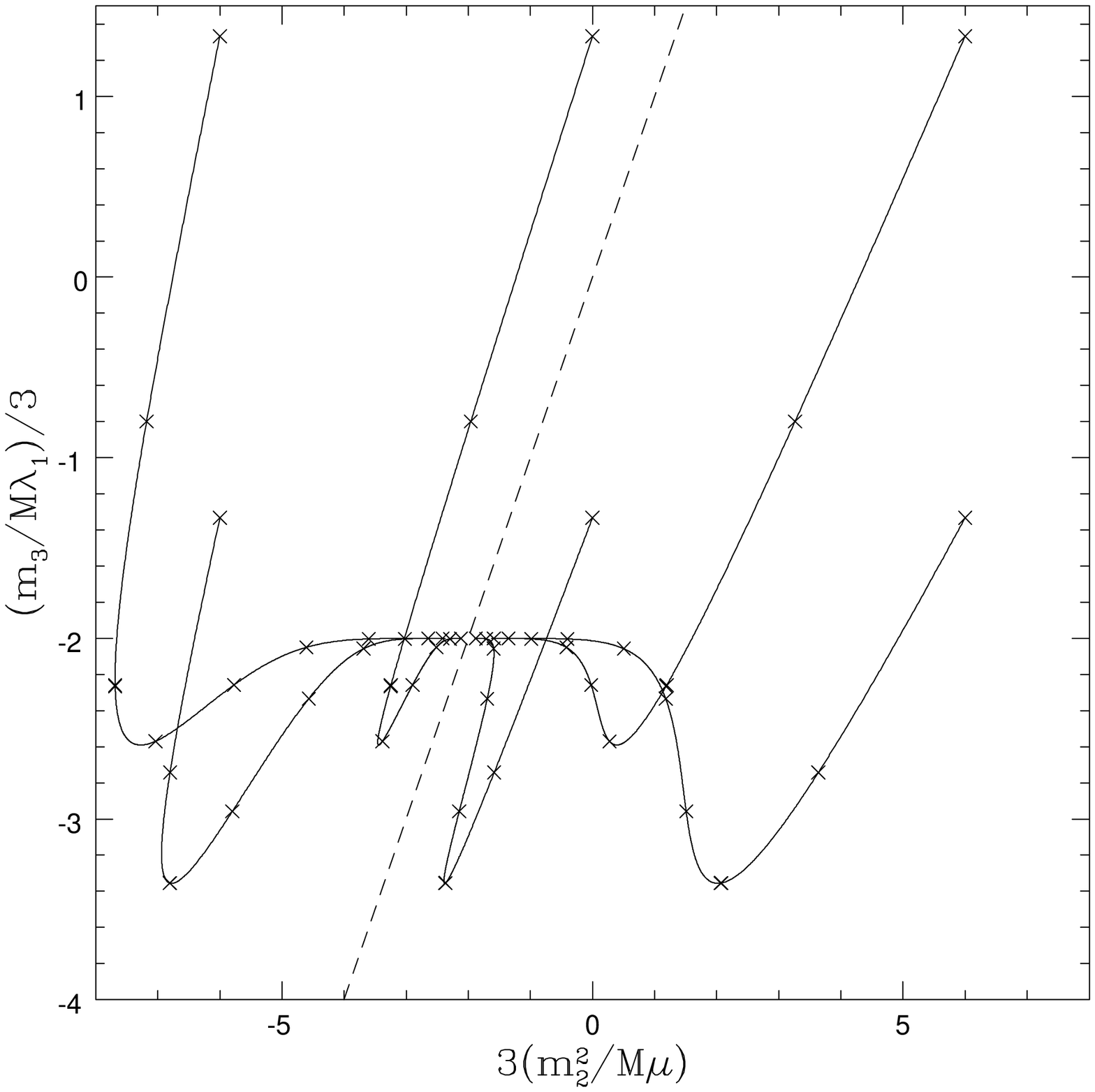}{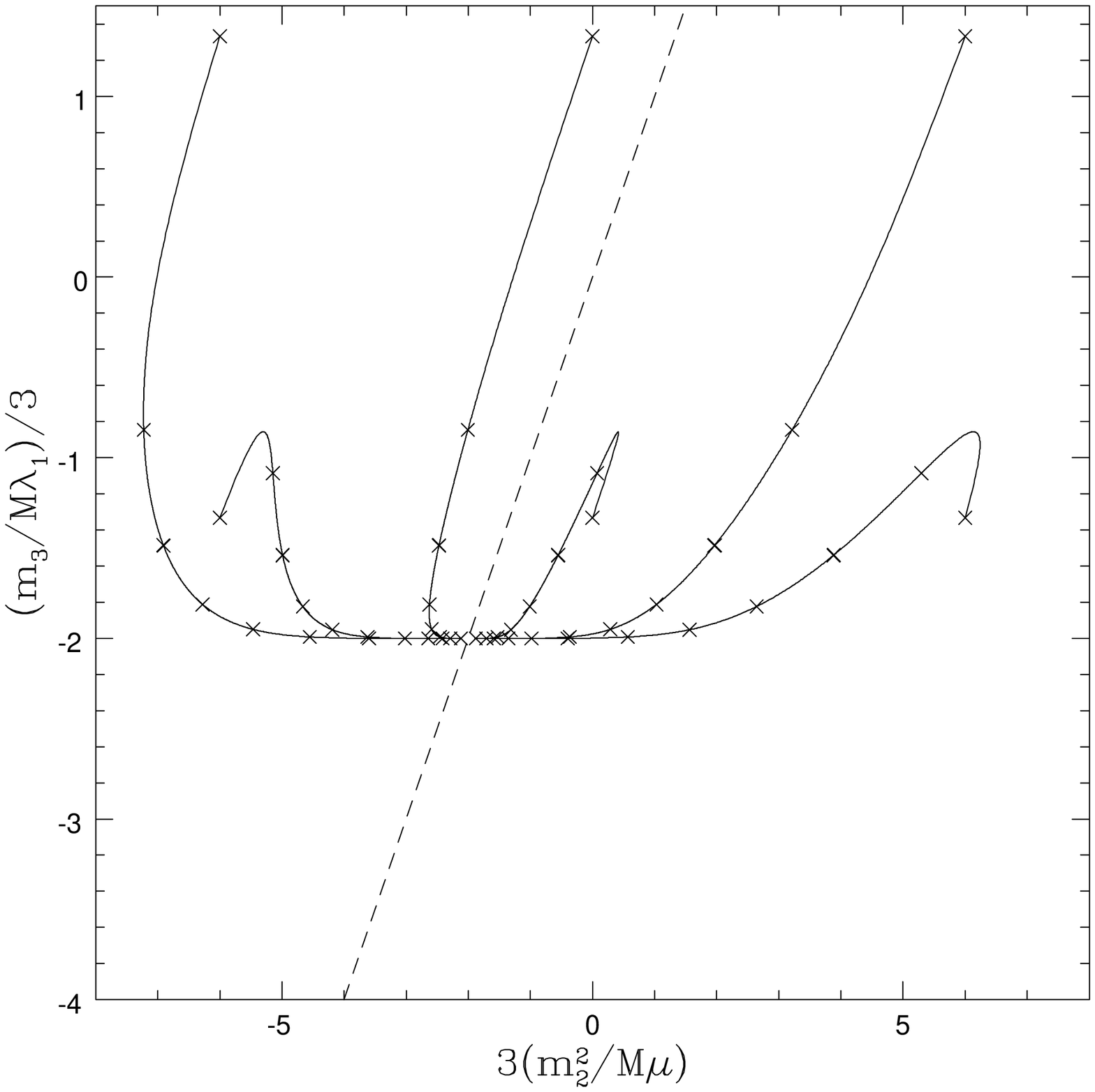}
\caption[]{\it
RG flow of the soft SUSY-breaking parameters in SUSY SU(5) with
$\beta=-8$ for initial conditions with
a) $\lambda_1(M^*) \! = \! 0.3$ 
and b) $\lambda_1(M^*) \! = \! 2.0$.
Every decrease of the
scale by a factor $10^{3/2}$ is marked on the flow.}
\label{a}
\end{figure}

\begin{figure}[ht]
\twographs{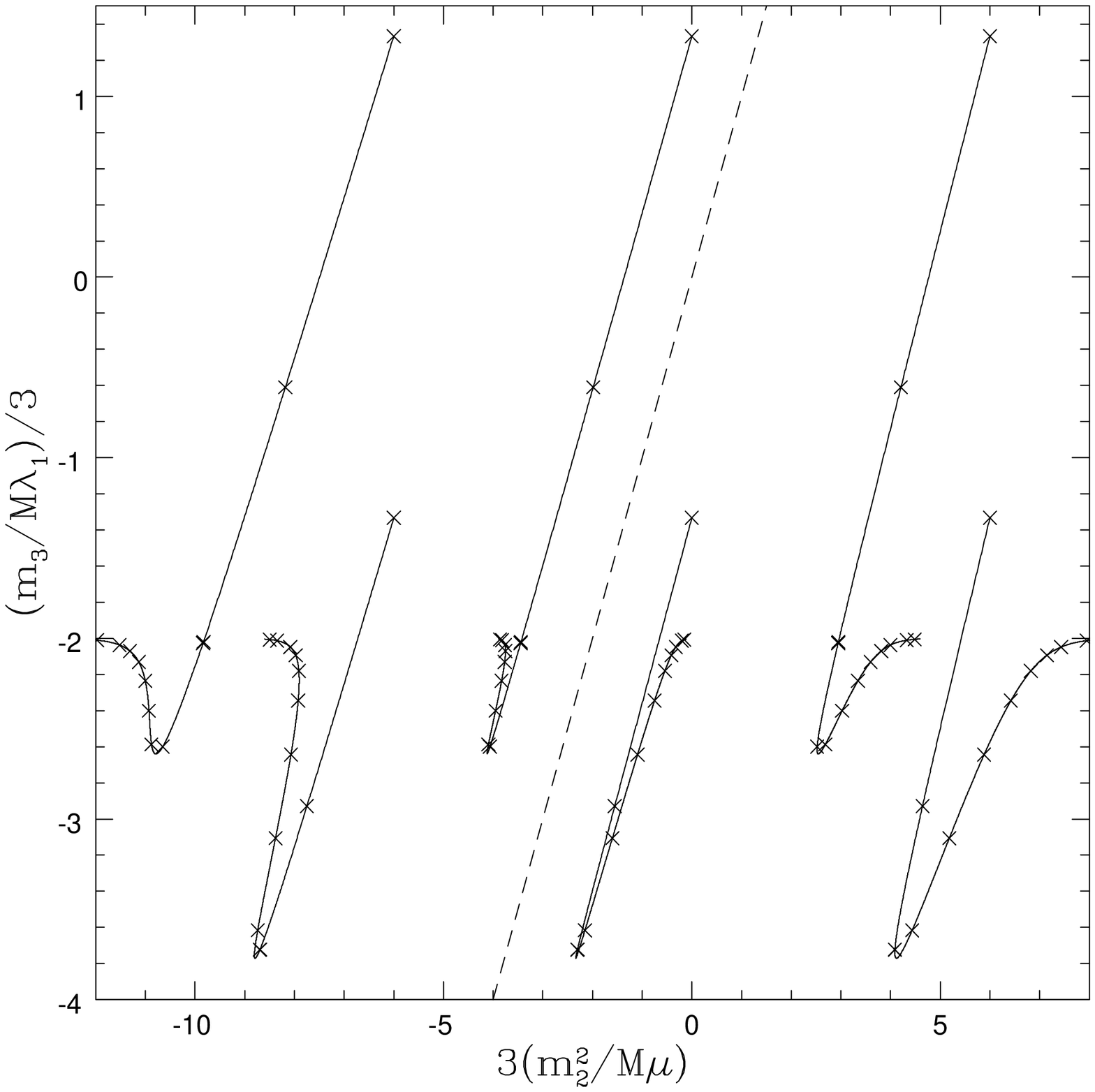}{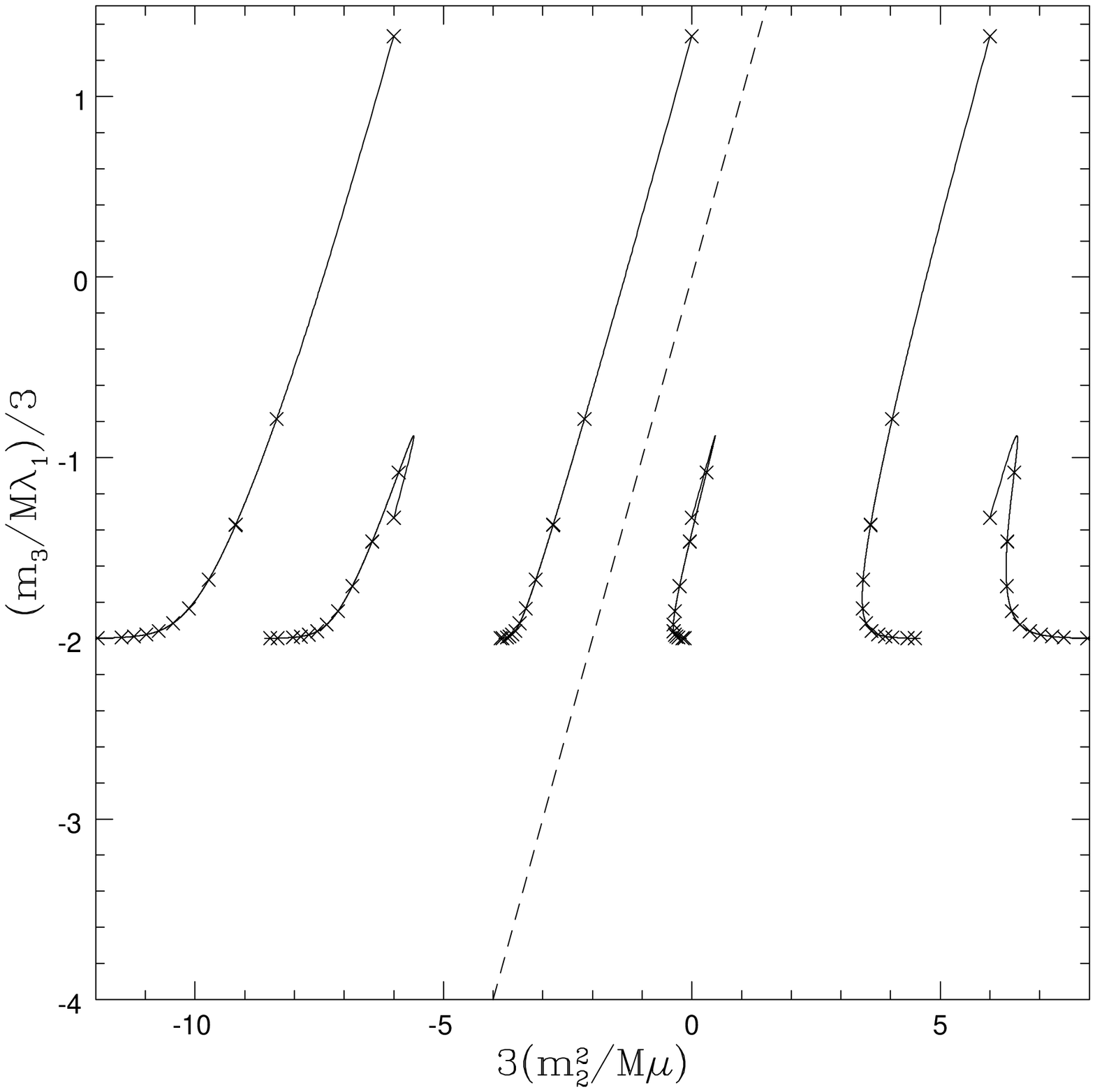}
\caption[]{\it  Same as Fig.\protect\ref{a}, but now for a model with
$\beta=2$.}
\label{b}
\end{figure}

We have also studied our 
RGEs numerically for the parameter values
$M^* \! = \! 10^{19}$~GeV,
$\mu \! = \! M_X \! = \! 10^{16}$~GeV,
$M(M^*) \! = \! m_{3/2} (M^*) \! = \! 10^{-13} M_X$
as advocated in most phenomenological soft SUSY-breaking scenarios.
The gauge coupling is fixed by $g^2 (M_X) \! = \! 8\pi/5$
to ensure consistency of SUSY SU(5) unification of the Standard Model
couplings with the low-energy values of the Standard Model couplings.
The running parameters $M$ and $\mu$ evolve
slowly, {\it i.e.} \/they decrease by 
a factor of $1/2$ between the Planck
and the GUT scale;
consequently their ratio 
in (\ref{fapp}) does not change sign. Hence, once the initial conditions
are fixed, the sign of the function $F$ depends on the relative 
magnitude of the combinations
of parameters $m_3/(3 M \lambda_1)$ and
$3 m_2^2 / (M \mu)$. 
The flow of these is depicted in Fig.~\ref{a} 
for a small and a large initial  
value of $\lambda_1 (M^*)$. 
The dashed line marks $3m_2^2/(M\mu)=m_3/(3M\lambda_1)$ where $F=0$.
The region to the left of this line corresponds to
the breaking of SU(5) to $G_{SM}$ while the region to the
right corresponds to unbroken SU(5).
For all the chosen initial values 
we have checked numerically that the
contribution of $\delta_{3/2}^2$ is indeed negligible
over the whole range of the running.
The figure clearly displays the attracting fixed point;
however, the attraction is typically rather weak 
between the Planck scale (first mark on the flow),
and the GUT scale (third mark on the flow).
Interestingly, flows starting on the left (right)
of the dashed line stay
on the left (right);
therefore the flows never cross
the boundary between
the region of parameter space corresponding
to unbroken SU(5)
and the one corresponding
to SU(5) breaking to $G_{SM}$.
This behaviour is also present
if the coefficient of the 
beta function is positive. For example, 
Fig.\ref{b} 
shows the flow diagram for the same initial
conditions as in Fig.\ref{a}, but now taking $\beta \! = \! 2$ in the RGEs 
(\ref{FP1})-(\ref{FP3}).
{}From Fig.\ref{b} it is clear that for $\beta \! >  \! 0$
a saddle point, rather than a fixed point, is present,
and the trajectories flow away from the dashed line.
This general property implies that
the running does not affect the amount
of tuning needed for the
phenomenologically desirable scenario
of SU(5) breaking to $G_{SM}$, 
in the sense that
the region of parameter space
supporting this scenario is mapped into itself by the RG flow.
We conclude that, 
while it does not require any fine tuning,
the scenario with SU(5) breaking to the Standard Model
is not a compelling prediction 
of the infra-red RG structure of SUSY SU(5).


We have limited ourselves to 
a zero-temperature analysis.
However, an important constraint on GUTs
is the consistency with a working cosmological scenario,
and checking this consistency requires in general
a finite-temperature analysis.
While we postpone this type of study to future work,
we would like to make some statements
concerning the possibility of cosmological implications 
of {\it Renormalisation Group Naturalness} 
\/analyses of the type here reported.

An important factor affecting 
{\it supercosmology} \cite{supercosmo,strongcoupling,longtime}
is the near degeneracy (up to SUSY breaking
terms) of several minima, which we mentioned above.
The free-energy difference between
the absolute minimum
and the other minima is of order SUSY breaking terms,
and therefore much smaller than the GUT scale.
In such cases one finds, at least within a perturbative analysis,
that even when the temperature becomes low enough
for the features of the zero-temperature effective potential
to be relevant,
the universe does not rapidly
reach the vacuum corresponding
to the absolute minimum of the
zero-temperature effective potential \cite{longtime}.
Actually, 
estimates within ordinary perturbative approaches
suggest that the time needed for the transition to the
true vacuum should be longer than the lifetime
of the universe \cite{longtime}.

One way to obtain working supercosmology
scenarios
is to advocate \cite{strongcoupling} 
thermal strong-coupling effects,
which are indeed at work in SUSY GUTs \cite{kapu}.
The investigation of these issues requires a careful
(and very delicate) thermal analysis which goes beyond the scope
of this paper. However, it should be noticed that
the type of analysis given here is not very relevant to
this type of supercosmological scenarios.

A more conventional, but {\it ad hoc}, way to
obtain working supercosmological scenarios
is based \cite{meltemp}
on fine tuning of the parameters of the Higgs potential.
One scales down the entire superpotential,
so that the height of the potential barrier between
competing vacua is of the same order 
as their energy difference,
while keeping fixed
the mass of the
gauge bosons mediating proton decay.
For example in the SUSY SU(5) GUT 
one would divide \cite{meltemp} 
both $\lambda_1$ and $\mu$ by a common 
large factor of order $10^{12}$, 
so that the ratio
$\mu / \lambda_1$ giving mass to the
gauge bosons mediating proton decay remains unchanged
\begin{eqnarray}
\lambda_1' \sim 10^{-12} \lambda_1 ~,~~~
\mu' \sim 10^{-12} \mu ~,
\label{scalingdown} \\
M_X' \sim {\lambda_1' \over \mu'} \sim
{\lambda_1 \over \mu} \sim M_X \,.~~~~~~~
\end{eqnarray}
Analyses of the type
advocated in the present paper
could be relevant for this supercosmology scenario;
one can in fact check the 
level of fine tuning at the Planck scale
needed to have, say,
a $10^{-12}$ fine tuning at the GUT scale.
We find that
the fine-tuned values of
$\lambda$ and $\mu$
are so far from the region of attraction
of the fixed point that the RG running between $M_P$ and $M_X$
is not substantial; {\it e.g.},
a fine tuning of $10^{-13}$ is
required at the Planck scale in order to obtain a $10^{-12}$ fine tuning
at the GUT scale.

The SUSY SU(5) GUT examined here is a toy model because, {\it e.g.},
it does not break electroweak symmetry. 
We believe that it would be interesting to investigate 
whether some of the issues exposed here
affect the analysis of phenomenologically relevant models.
If a non-trivial fixed point structure was found also in those more
complicated models it could have important implications
for the associated SSB physics. 
Similarly, 
there might be important implications
if it was found that even
in phenomenologically relevant models
certain forms of the Higgs potential
discussed at tree level in the recent literature are not 
renormalisation group invariant.

\section*{Acknowledgments}
We have greatly benefited from conversations with
G.~Ross, which we happily acknowledge. We also thank G.~Ross for
comments on the letter.
B.A. would also like to thank 
M. Drees and S. Pallua
for correspondence.
G.A.-C. would like to thank
R. Barbieri, F. Buccella,
D. Nanopoulos, K. Tamvakis,
and C. Wetterich for useful discussions.
The work of G.A.-C. was supported by the EC Human and Capital Mobility
program and by PPARC\@.
The work of O.P. was 
supported in part by the EC Human and Capital Mobility program.

\end{document}